\documentclass[aps,prl,twocolumn,showpacs,superscriptaddress,nofootinbib,10pt]{revtex4-2}

\usepackage{acronym}
\usepackage{amsfonts}
\usepackage{amsmath}
\usepackage{amssymb}
\usepackage{bm}
\usepackage{booktabs}
\usepackage{cases}
\usepackage{color}
\usepackage{comment}
\usepackage[dvipsnames]{xcolor}
\usepackage{enumerate}
\usepackage{epsfig}
\usepackage{etoolbox}
\usepackage{fancyref}
\usepackage{fancyhdr}
\usepackage[T1]{fontenc} 
\usepackage{graphicx}
\usepackage{hyperref}
\usepackage{indentfirst}
\usepackage{latexsym}
\usepackage{mathrsfs}
\usepackage{mciteplus}
\usepackage{multirow}
\usepackage{rotating}
\usepackage{scrextend}
\usepackage{soul}
\usepackage{subfigure}
\usepackage{verbatim}

\newcommand{\DeltaASB}{\Delta^{}_\mathrm{ASB}}
\newcommand{\DeltaESB}{\Delta^{}_\mathrm{ESB}}

\def\BAAS{Bull. Am. Astron. Soc.}
\def\CQG{Classical Quantum Gravity}
\def\CMP{Commun. Math. Phys.}

\def\GRG{Gen. Relativ. Gravit.}

\def\JHEP{JHEP}
\def\JMP{J. Math. Phys.}

\def\LRR{Living Reviews in Relativity}

\def\NCB{Nuovo Cimento B}

\def\PRD{Phys. Rev. D}                               
\def\PRL{Phys. Rev. Lett.}

\def\RPP{Rept. Prog. Phys.}
\def\RMP{Rev. Mod. Phys.}

\begin{document}

\title[Probing fuzzballs and beyond: Model-independent tests of Kerr symmetry breaking with LISA]{Probing fuzzballs and beyond: Model-independent tests of Kerr symmetry breaking with LISA}

\author{Pablo F. Muguruza}
\email{pfernandez@ice.csic.es}
\affiliation{Institut de Ci\`encies de l'Espai (ICE, CSIC), Campus UAB, Carrer de Can Magrans s/n, 08193 Cerdanyola del Vall\`es, Spain}
\affiliation{Institut d'Estudis Espacials de Catalunya (IEEC), Edifici Nexus, Carrer del Gran Capit\`a 2-4, despatx 201, 08034 Barcelona, Spain}
\affiliation{Department of Physics, Universitat Autònoma de Barcelona, Facultat de Ci\`encies, Edifici C, 08193 Bellaterra (Cerdanyola del Vallès), Spain}

\author{Carlos F. Sopuerta}
\email{carlos.f.sopuerta@csic.es}
\affiliation{Institut de Ci\`encies de l'Espai (ICE, CSIC), Campus UAB, Carrer de Can Magrans s/n, 08193 Cerdanyola del Vall\`es, Spain}
\affiliation{Institut d'Estudis Espacials de Catalunya (IEEC), Edifici Nexus, Carrer del Gran Capit\`a 2-4, despatx 201, 08034 Barcelona, Spain}

\begin{abstract}
Gravitational waves provide a unique probe of the strong‑field regime of gravity, offering access to physics beyond the classical black hole paradigm. We explore how space‑based observations of extreme-mass-ratio inspirals (EMRIs) by the Laser Interferometer Space Antenna (LISA) can be used to test for departures from the Kerr geometry, and in particular for the breaking of its axial and equatorial symmetries, as expected in quantum-gravity-inspired alternatives to Kerr black holes such as the fuzzball proposal. By introducing generic multipolar deformations encoding potential symmetry breakings and performing a systematic parameter estimation analysis, we forecast LISA's ability to constrain such departures from the Kerr geometry in the strong-field regime. We show that EMRI signals with realistic signal-to-noise ratios can constrain several higher-order multipoles of the primary, opening a new observational window onto strong-field deviations from the Kerr geometry. In particular, we find that LISA can constrain generic nonaxisymmetric mass quadrupole deformations at the $10^{-4}-10^{-2}$ level and axisymmetric mass octupole deformations at the $10^{-1}$--$10^{0}$ level, providing concrete observational targets for strong-field deviations from the Kerr geometry, such as those that may arise in fuzzball and other exotic-compact-object scenarios. Our results demonstrate that precision measurements of EMRI waveforms will transform LISA into a powerful laboratory for fundamental physics and offer precise constraints on quantum-gravity-motivated models of compact objects.
\end{abstract}

\maketitle

\noindent\textbf{Introduction}.
One of the main challenges in theoretical physics is the construction of a consistent quantum theory of gravity that reconciles the principles of quantum mechanics with those of general relativity (GR).  
Among the many attempts to get into the realm of quantum gravity we have string theory,
which provides a concrete framework to address this problem. It also offers a microscopic description of black hole (BH) thermodynamics that resulted in the derivation by Strominger and Vafa~\cite{Strominger:1996sh} of the Bekenstein-Hawking entropy~\cite{Bekenstein:1973ur,Hawking:1974rv} by counting, for certain supersymmetric black holes in string theory, the degeneracy of underlying quantum states.

Following this result, considerable effort has been devoted to understanding the spacetime realization of individual BH microstates. Rather than appearing as small quantum corrections localized near a classical horizon, it has been found that large classes of microstates admit smooth, horizonless geometries that reproduce the same asymptotic charges as the corresponding classical BH solutions while differing from them at horizon scales~\cite{Skenderis:2006ah,Skenderis:2008qn,Lunin:2001jy,Lunin:2002qf}. 
These results led to the \emph{fuzzball} paradigm~\cite{Mathur:2005zp}, in which individual BH microstates are described by distinct horizonless configurations, and the classical BH geometry with an event horizon emerges only as a coarse-grained thermodynamic description of an ensemble of such states~\cite{Mathur:2005zp,Skenderis:2008qn}. 
That is, in the fuzzball picture, the notion of event horizon is replaced by a quantum region with structure extending over scales comparable to the would-be horizon radius. 
Hence, deviations from the classical Kerr geometry may persist at horizon scales and can potentially affect strong-field observables.
The fuzzball model has also been advocated as a concrete resolution of the BH information problem, as it removes the need for information loss behind an event horizon while remaining consistent with BH thermodynamics at the coarse-grained level~\cite{Mathur:2005zp,Chowdhury:2007jx,Mathur:2008kg,Chowdhury:2008bd}.

Then, the fuzzball model offers the possibility that sufficiently precise probes of strong-field gravity could be sensitive to quantum-gravitational structure, even for astrophysical-size objects.  
In this context, gravitational wave (GW) observations provide an alternative to particle colliders for testing aspects of string theory.
In particular, GW observations of binary BH coalescences provide a unique and promising avenue for such tests, as they directly encode the nonlinear dynamics of spacetime in the radiative sector of the strong-field regime.

The current second generation of ground-based GW detectors, LIGO, Virgo and KAGRA~\cite{Brady2020}, has already carried out tests of GR and fundamental physics that include~\cite{LIGOScientific:2016lio,LIGOScientific:2019fpa,LIGOScientific:2020tif,LIGOScientific:2021sio,LIGOScientific:2025rid,LIGOScientific:2025wao}: Inspiral-merger-ringdown consistency tests, parametrized deviations from post-Newtonian dynamics, constraints on modified dispersion relations, bounds on additional GW polarization states, and limits on the graviton Compton wavelength. 
However, their current sensitivity is not enough to probe horizon-scale quantum structure of the type predicted by the fuzzball paradigm, which motivates the need for the next generation of GW detectors~\cite{Berti:2015itd,Barack:2018yly}. 
On the ground, third-generation detectors such as the Einstein Telescope~\cite{Punturo:2010zz} and Cosmic Explorer~\cite{Reitze:2019iox} are expected to improve the GW strain sensitivity by approximately one order of magnitude, allowing for a much broader science case including fundamental physics~\cite{Sathyaprakash:2012jk,ET:2019dnz,ET:2025xjr}. 
In space, the Laser Interferometer Space Antenna (LISA)~\cite{LISA:2017pwj} will access the millihertz GW band, between $10^{-4}\,$Hz and $1\,$Hz, enabling a rich and complementary science program~\cite{Colpi:2024xhw,LISA:2022yao,LISACosmologyWorkingGroup:2022jok}, targeting massive BH binaries and extreme-mass-ratio inspirals (EMRIs), LISA is expected to revolutionize our knowledge of BHs and the gravitational interaction~\cite{Barausse:2020rsu,LISA:2022kgy}.

In this work, we advocate for the use of LISA GW observations of EMRIs~\cite{Babak:2017tow,Berry:2019wgg} as a tool for testing the fuzzball paradigm. EMRIs are binary systems consisting of a primary with mass $M\in[10^{5},10^{7}],M_\odot$, the putative massive BH (with mass $M\in[10^5,10^7]M_\odot$), and a secondary compact object, typically a neutron star or a stellar-mass BH (with mass $m\sim[1,10^2]\,M_\odot$) orbiting the primary along a highly relativistic inspiral trajectory.  
Due to the extreme mass ratios involved, $q\equiv m/M\ll 1$ (typically $10^{-6} < q < 10^{-4}$), EMRIs are long-lived GW sources that can emit, within the LISA frequency band, of the order of $N\sim q^{-1}\sim 10^{4-6}$ cycles of a highly phase-coherent GW signal that encodes very precise information about the geometry of the primary, effectively providing an exquisite probe of its multipolar structure. 
We then expect that EMRI observations with LISA will give access to the strong-field regime around massive compact objects~\cite{Berry:2019wgg,Cardenas-Avendano:2024}, making them a powerful laboratory for testing exotic compact objects, such as those of the fuzzball paradigm, and more generally for strong-field deviations from the classical BH description predicted by GR~\cite{Hawking:1973uf}.

~

\noindent\textbf{Framework to test fuzzballs with EMRIs}.  The exterior geometry of a stationary and axisymmetric compact object in vacuum GR is uniquely characterized by its gravitational multipole moments defined at spatial infinity~\cite{Geroch:1970rg,Hansen:1974ro,Thorne:1980rm}, the mass and current moments: $\{\mathcal{M}_\ell,\mathcal{S}_\ell\}_{\ell=0,\ldots,\infty}$. For a Kerr BH~\cite{Kerr:1963ud}, the BH uniqueness (\emph{no hair}) theorems~\cite{Israel:1966rt,Carter:1971zc,Hawking:1972} imply that all these multipole moments are completely determined by the BH mass $M$ and magnitude of its spin angular momentum $|\mathbf{S}|$ (with the spin parameter satisfying: $|a| \equiv|\mathbf{S}|/M\leq M$). They follow the remarkably simple closed-form relations~\cite{Hansen:1974ro}:
\begin{equation}
\mathcal{M}^{}_{\ell}+i\mathcal{S}^{}_{\ell}= M (ia)^{\ell}\,.
\label{Kerr-relations}
\end{equation}

In general, fuzzball microstates need not satisfy the symmetries of the Kerr family (see~\cite{Mathur:2005zp,Skenderis:2007yb,Skenderis:2008qn}) and in general they can break both axisymmetry and equatorial symmetry, leading to a multipolar structure that is not constrained by the Kerr relations (see~\cite{Cardoso:2016oxy,Cardoso:2019rvt} for accounts on exotic compact objects and their consequences).
Observational constraints on fuzzballs have been discussed in~\cite{Mayerson:2020tpn,Bianchi:2020bxa,Bena:2020uup,Bacchini:2021fig,Mayerson:2023wck}. 

However, the analysis presented here is not exclusive to fuzzballs. Any horizonless or exotic compact object, or even a beyond-GR black hole, whose multipole moments deviate from the Kerr values in the way we parametrize would produce the same observational signatures. Our use of the fuzzball paradigm is therefore representative of a broad class of near-horizon modifications of the Kerr geometry rather than a unique prediction of it, and the bounds we derive are, strictly, constraints on Kerr-symmetry-breaking multipoles. Is important to remark that our analysis does not involve any specific fuzzball model. The deformation parameters are treated agnostically as proxies for such deviations, and no quantitative mapping between a specific microstate solution and the constrained moments is assumed.
We also note that whether generic fuzzball microstates possess multipole moments observably different from Kerr is under debate. Typical microstates are expected to be observationally indistinguishable from the corresponding black hole, and only atypical or diffuse configurations would give multipole deviations large enough to be detectable~\cite{Raju:2018xue}. Our forecasts should therefore be understood as model-independent constraints on Kerr-symmetry breaking.

To determine the extent to which a space-based observatory such as LISA can constrain deviations from the standard EMRI scenario, i.e. a MBH $+$ NS/BH in GR, we need to build an EMRI waveform model in which the multipolar structure of the primary is extended beyond the Kerr metric. 
That is, an EMRI model where a number of multipole moments are promoted to independent physical parameters so that the model can encompass fuzzballs as well as other exotic compact objects of interest, such as boson stars or gravastars. 
In practice, this means we have to modify both the orbital dynamics of the secondary as well as the inspiral dynamics driven by gravitational radiation reaction due to the time-dependent EMRI spacetime geometry. 
However, the radiation-reaction timescale is much longer than the orbital timescale, $T_{\mathrm{rr}} \sim q^{-1} T_{\mathrm{orb}} \gg T_{\mathrm{orb}}$, which allows for a separate treatment.  
Finally, we need a prescription to evaluate the EMRI waveforms given the EMRI dynamics and use them to derive projected constraints on deviations from the Kerr multipolar structure. This program was initiated by Ryan~\cite{Ryan:1995wh,Ryan:1995zm,Ryan:1995xi,Ryan:1997hg,Ryan:1997kh}, who demonstrated its feasibility and showed that the \emph{classic} LISA design~\cite{Danzmann:1996da,Cutler:1998rf,Prince:2002hp} could allow for the measurement of three to five multipole moments, corresponding to one to three independent tests of the Kerr hypothesis~\cite{Ryan:1997hg}. 

Later, Barack and Cutler, using their \emph{analytic Kludge} (AK) EMRI model~\cite{Barack:2003fp}, which contains all the main ingredients of the EMRI dynamics in a Newtonian and post-Newtonian fashion, were able to introduce the mass quadrupole, $\mathcal{M}^{}_2$ ($= -Ma^2 = -S^2/M$ for a Kerr BH), as an additional independent parameter~\cite{Barack:2006pq}. Recent studies have shown~\cite{Babak:2017tow} that LISA can constrain fractional deviations away from the Kerr mass quadrupole $Q$ at the level of $\sim 10^{-4}-10^{-2}$. 
Fransen and Mayerson~\cite{Fransen:2022jtw} went beyond by including a current quadrupole and an axisymmetric mass octupole moment as independent parameters. The contribution from these parity-odd moments breaks equatorial symmetry, which is one of the two fundamental symmetries of the Kerr metric. They found that LISA could constrain such symmetry-breaking effects at the level of $10^{-2}$~\cite{Fransen:2022jtw}. 

In this work, we incorporate the \emph{axisymmetric and nonaxisymmetric components of the mass quadrupole and octupole}, thereby constraining the two fundamental symmetries of the Kerr spacetime (equatorial and axial symmetry) within a single framework and providing a new observational avenue to test fuzzballs with LISA EMRI observations. 
Whereas previous analyses treated equatorial-symmetry breaking~\cite{Fransen:2022jtw} and axisymmetry breaking~\cite{Loutrel:2022ant,Loutrel_2024} separately, we combine both within a single theory-agnostic EMRI model and, for the first time, compute the dissipative (radiation-reaction) contributions of these axisymmetric and nonaxisymmetric quadrupole and octupole moments to the secular inspiral. This provides a joint LISA forecast on the breaking of both Kerr symmetries, of direct relevance to fuzzballs and other Kerr-symmetry-breaking compact objects.

~

\noindent\textbf{20-parameter EMRI model for testing the fuzzball paradigm}.
Probing deviations from the Kerr spacetime does not require to include arbitrarily high multipole moments. 
Already at the quadrupole-octupole level, nonaxisymmetric components of the mass quadrupole, such as polar or axial configurations, we can place stringent constraints on non-Kerr geometries~\cite{Loutrel:2022ant, Loutrel_2024}, in particular in complex effective geometries like the ones expected from fuzzballs. 
Here, we go beyond the work of~\cite{Fransen:2022jtw} by incorporating all possible symmetry breakings at the quadrupolar and octupolar level within a theory-agnostic single model. 
We parametrize these deviations from the Kerr geometry through the nonaxisymmetric mass quadrupole and axisymmetric mass octupole. 
These moments serve as proxies to quantify LISA's sensitivity to violations of axisymmetry and equatorial symmetry, respectively. 

In principle, one could also introduce the current quadrupole to probe violations of equatorial symmetry. 
However, it was shown in~\cite{Fransen:2022jtw} that its inclusion affects the results only at the level of $\lesssim 10\%$, and we therefore neglect it here. 
We could also include the polar nonaxisymmetric octupole moment, which captures simultaneously violations of the axial and equatorial symmetries. 
However, since the nonaxisymmetric quadrupole and the axisymmetric octupole already provide independent and more stringent constraints on these symmetry breakings, the inclusion of this additional moment does not significantly improve the parameter-estimation accuracy. For this reason, we do not include it in our EMRI model.

Taking all these considerations into account, we construct a semianalytic EMRI waveform model that provides a computationally efficient description of the inspiral while allowing for a generic multipolar structure of the EMRI primary.
While highly accurate EMRI waveforms are being developed within the gravitational self-force program~\cite{Poisson:2011nh,Barack:2018yvs}, as well as through adiabatic and postadiabatic reductions (see. e.g.~\cite{Hughes:2021exa,Katz:2021yft,Speri:2023jte}), such models are difficult to generalize to non-Kerr geometries. 
Our approach instead prioritizes flexibility, enabling a systematic exploration of deviations in the multipolar structure of the central compact object, the EMRI primary.  
As in the AK model~\cite{Barack:2003fp}, the underlying orbital dynamics of our model is Newtonian.
However, rather than introducing post-Newtonian corrections in a   phenomenological way, we derive  the orbital dynamics and the radiation-reaction effects consistently from the multipolar gravitational potential and the quadrupolar emission approximation, respectively (see~\cite{Muguruza:2026hqn} for details).  
The resulting model defines a 20-dimensional parameter space. We neglect the spin of the secondary compact object, but aside from this simplification, the orbital dynamics is completely generic. 
The complete set of parameters is summarized in Table~\ref{tab:parameterspace}, where we distinguish between intrinsic EMRI properties, extrinsic parameters describing the source location and orientation, and a set of multipolar parameters characterizing the primary.

The orbital dynamics is described within an effective one-body framework at leading order in the mass ratio. The conserved quantities associated with the orbital motion are the energy ($E$), arising from time-translation invariance, and the $z$-component of the orbital angular momentum ($L_z$), arising from axial symmetry. In spherical coordinates $(r,\theta,\phi)$, they are given by: 
\begin{widetext}
\begin{eqnarray}
E & = &\frac{1}{2} \mu (\dot{r}^2 + r^2 \dot{\theta}^2 + r^2 \sin^2\theta \dot{\phi}^2) \nonumber \\[2mm] 
& - &\frac{M}{r} + \frac{M Q}{2r^3} (1 - 3\cos^2\theta) - \frac{3 M Q^{}_{+}}{2r^3} \sin^2\theta \cos[2 \varphi^{}_Q] - \frac{M \mathcal{O}}{2 r^4} (5\cos^3\theta - 3\cos\theta) - \frac{15 M \mathcal{O}^{}_{+}}{2 r^4} \sin^2\theta \cos\theta \cos[3 \varphi_{\mathcal{O}}]\,, \label{energy} 
\end{eqnarray}
\end{widetext}
and 
\begin{equation} \label{orbitalmomentumz}
L^{}_z = r^2 \sin^2\theta \dot{\phi} - \frac{2 S \sin^2\theta}{r}\,,
\end{equation}
where $S$ denotes the spin of the EMRI primary (aligned with the $z$-axis), $Q$ and $Q_{+}$ denote the axisymmetric and nonaxisymmetric quadrupole moments, respectively, while $\mathcal{O}$ and $\mathcal{O}_{+}$ are the corresponding octupole moments. The angles $\varphi_Q$ and $\varphi_{\mathcal{O}}$  parametrize the orientation of the nonaxisymmetric deformations. 
The orbital equations of motion are derived from the associated Lagrangian and expressed in terms of the standard orbital elements describing eccentric and inclined motion. The multipolar structure of the central object modifies the effective potential, inducing corrections to orbital frequencies and  precession effects.

The inspiral evolution is driven by gravitational radiation reaction. We compute the energy and angular-momentum fluxes using the quadrupole formula for gravitational-wave emission (see, e.g.,~\cite{Maggiore:2007mm}):
\begin{eqnarray}
\frac{dE}{dt} & =  & \frac{1}{5} \left\langle \dddot{M}^{}_{ij}\dddot{M}^{}_{ij}
- \frac{1}{3} (\dddot{M}_{kk})^2 \right\rangle\,, 
\label{energy-gw-flux} \\[2mm]
\frac{d L^i}{dt} & = &  -\frac{2}{5}\,\epsilon^{ikl} \left\langle \ddot{M}^{}_{kj}\dddot{M}^{}_{lj}  \right\rangle\,,
\end{eqnarray}
with $\epsilon^{ikl}$ the three-dimensional Levi-Civita symbol. Using the solution of the orbital motion, we express these fluxes in terms of the orbital variables and expanding perturbatively in the post-Newtonian parameter $\epsilon^2=M/r$. We then average these fluxes over an orbital period to obtain the secular evolution equations for the orbital parameters. In this way, our EMRI model incorporates, for the first time within this framework, the dissipative contributions of both axisymmetric and nonaxisymmetric quadrupole and octupole moments to the secular evolution of the orbital frequency and eccentricity (see~\cite{Muguruza:2026hqn} for details).

~

\noindent\textbf{Forecasts for LISA constraints on fuzzballs}. The waveform is generated by first integrating the orbital evolution equations backward in time from the last stable orbit (LSO), including the secular evolution of the orbital elements $(\nu,e,\tilde{\gamma},\alpha)$ under gravitational radiation reaction.  The orbital inclination $\lambda$ is kept fixed, as its secular variation is negligible over the inspiral timescale.

Then, at leading quadrupolar order, the far-zone metric perturbation in the transverse--traceless (TT) gauge (see, e.g.~\cite{Misner:1973cw,Maggiore:2007mm}) is
\begin{equation}\label{eq:hijTT}
h^{\rm TT}_{ij}(t) = \frac{2}{D_L}
\left(P^{}_{ik}P^{}_{jl}-\frac{1}{2}P^{}_{ij}P^{}_{kl}\right)
\ddot{\mathcal{I}}^{}_{kl}(t_{\rm ret}) \,,
\end{equation}
where $D_L$ is the luminosity distance to the source, $t_{\rm ret}=t-D_L$ is the retarded time, and $P_{ij}=\delta_{ij}-\hat n_i\hat n_j$ is the projector orthogonal to the unit vector $\hat n$ pointing from the detector to the source. The tensor $\mathcal{I}_{kl}$ denotes the radiative mass quadrupole moment defined, which at leading order is given by $\mathcal{I}_{ij} = \mu x_i x_j$. 

The detector response is obtained by projecting the TT metric perturbation onto the LISA constellation using the standard long-wavelength approximation, including the time-dependent modulation induced by the orbital motion of the detector~\cite{Cutler:1998rf,Cornish:2003d} (see also~\cite{Barack:2003fp} for the implementation in the AK model). For the signal analysis, we employ the orthogonal time-delay interferometry (TDI) channels of LISA and adopt the corresponding instrumental noise model (see Appendix B in~\cite{Muguruza:2026hqn} for details).

Parameter uncertainties have been estimated through a Fisher matrix analysis. Because the Fisher matrix is often ill-conditioned due to strong correlations among parameters, as is the case here, we compute its pseudoinverse using singular-value decomposition. A small cutoff is applied to the singular values in order to regularize the inversion and ensure numerical stability. We also performed a series of consistency checks (see~\cite{Muguruza:2026hqn} for details) to validate both the correctness of our dynamical equations and the robustness of the matrix inversion procedure. In particular, we have checked that for standard EMRIs we get equivalent forecasts as in other studies~\cite{Babak:2017tow}.

\begin{table}[t]
\caption{Parameters of our 20-dimensional EMRI model.}
\centering
\small
\renewcommand{\arraystretch}{1.05}
\begin{tabular}{@{} l p{7cm} @{}}
\toprule
Symbol & Description \\
\midrule

\multicolumn{2}{c}{\textbf{Intrinsic parameters}} \\
\midrule
$M$ & Mass of the primary  \\
$\mu$ & Mass of the secondary \\
$\tilde{S}= a/M$ & Dimensionless spin magnitude of the primary  \\
$e^{}_\mathrm{LSO}$ & Orbital eccentricity at the LSO \\
$\cos\lambda$ & Cosine of orbital inclination ($\hat L\!\cdot\!\hat S$) \\
$\tilde\gamma^{}_\mathrm{LSO}$ & Pericenter angle at the LSO \\
$\Phi^{}_\mathrm{LSO}$ & Mean orbital anomaly at the LSO \\
\midrule
\multicolumn{2}{c}{\textbf{Extrinsic parameters}} \\
\midrule
$t^{}_\mathrm{LSO}$ & Time of LSO crossing (scaled to $1\,$mHz) \\
$\theta^{}_S$ & Polar angle of the EMRI sky location  \\
$\phi^{}_S$ & Azimuthal angle of the EMRI sky location \\
$\theta^{}_K$ & Polar Angle of the primary Spin orientation  \\
$\phi^{}_K$ & Azimuthal Angle of the primary Spin orientation \\
$\alpha^{}_\mathrm{LSO}$ & Azimuth of the angular momentum $\hat{L}$ at the LSO \\
$D^{}_L$ & Luminosity Distance to the EMRI system \\
\midrule
\multicolumn{2}{c}{\textbf{Dimensionless Multipole deviation parameters}} \\
\midrule
$\tilde Q$ & Spheroidal quadrupole moment \\
$\tilde Q^{}_+$ & Polar nonaxisymmetric quadrupole moment \\
$\psi^{}_Q$ & Argument of the nonaxisymmetric quadrupole \\
$\tilde{\mathcal O}$ & Spheroidal octupole moment \\
$\tilde{\mathcal O}^{}_+$ & Polar nonaxisymmetric octupole moment \\
$\psi^{}_{\mathcal O}$ & Argument of the nonaxisymmetric octupole \\
\bottomrule
\end{tabular}
\label{tab:parameterspace}
\end{table} 

\begin{figure}
    \centering
    \includegraphics[width=1.0\linewidth]{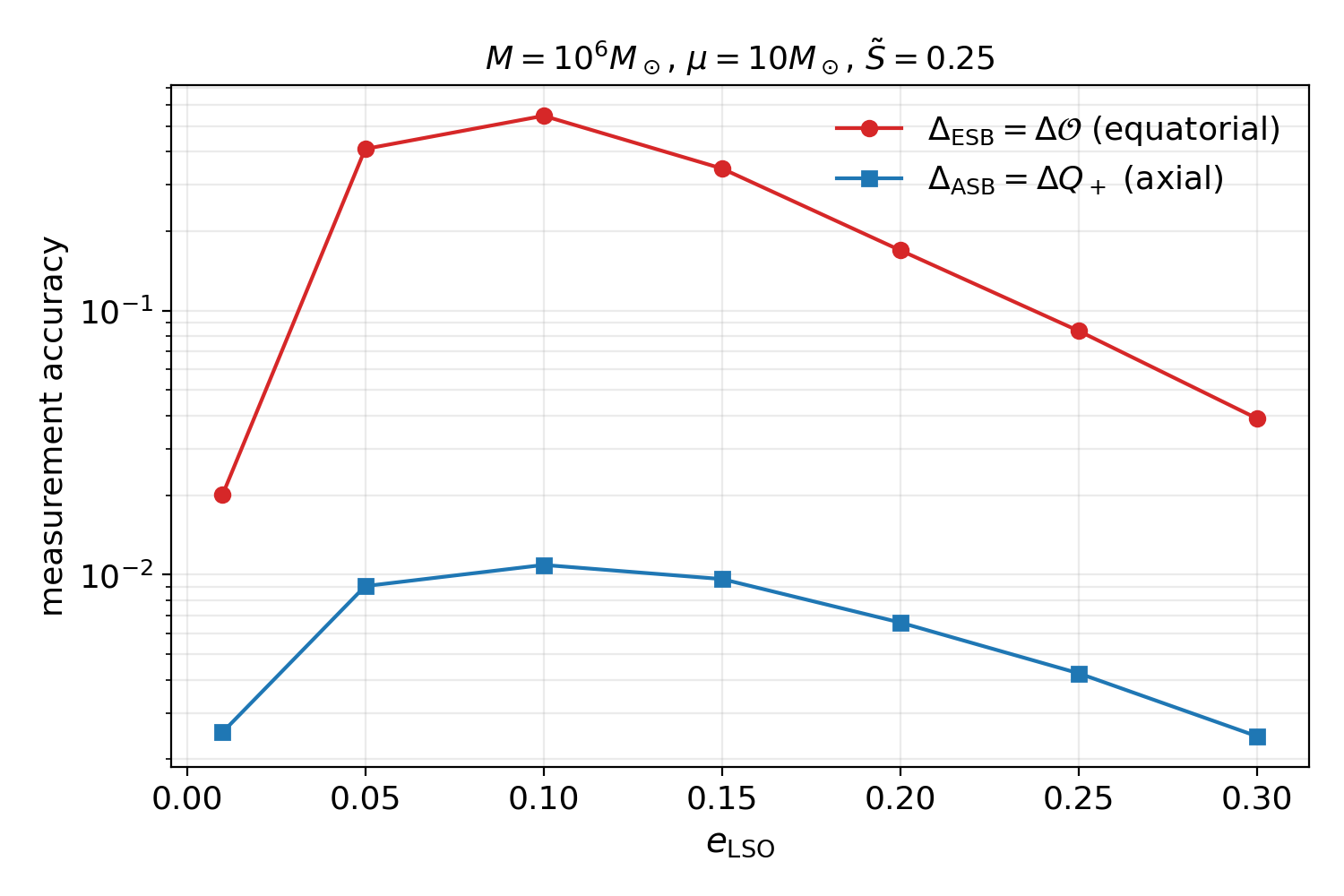}
    \caption{Constraints on the breaking of the two fundamental symmetries of the Kerr metric, equatorial and axial symmetry, shown as a function of the eccentricity $e_\mathrm{LSO}$, for a LISA EMRI observation. The central object has mass $M = 10^6 M_{\odot}$ and spin $\tilde{S} = 0.25$, while the secondary has mass $\mu = 10 M_{\odot}$. The angular parameters are fixed according to Table \ref{tablesconstraints}.}
    \label{fig:q_vs_e}
\end{figure}

For the purposes of testing the fuzzball paradigm, we focus on the estimation of the nonaxisymmetric quadrupole moment, associated with the breaking of axial symmetry ($\DeltaASB\equiv\Delta Q_+$), and of the axisymmetric component of the octupole moment, associated with the breaking of equatorial symmetry ($\DeltaESB\equiv\Delta \mathcal{O}$). Since only the loudest EMRI events are expected to be useful for such tests, we consider one-year-long EMRI signals, which is a conservative assumption. We adopt the LISA noise curve from~\cite{Bayle2022LISANode}. As a representative system, we consider an EMRI consisting of a secondary of mass $\mu=10\,M_{\odot}$ inspiralling into a primary of mass $M=10^{6}\,M_{\odot}$ and dimensionless spin $\tilde{S} = S/M^2 = a/M = 0.25$, with eccentricity at the LSO $e_\mathrm{LSO} = 0.1$. The typical signal-to-noise ratio (SNR) associated with this type of EMRI event is $\sim 30$. Then, we find we can constrain ESB at the $5.5\times10^{-1}$ level, while ASB can be constrained at the $1.1\times10^{-2}$ level, as illustrated in Fig.~\ref{fig:q_vs_e}. These results indicate that LISA will constrain ASB with substantially higher precision than ESB, with the achievable accuracy differing by more than one order of magnitude (a factor $\sim50$).

\begin{table}[ht]
\caption{LISA measurement accuracies for equatorial ($\DeltaESB$) and axial ($\DeltaASB$) symmetry-breaking parameters from EMRI observations, quantifying deviations from the Kerr geometry.
We have performed simulations for three EMRI primary masses, $M = 10^{5},\,10^{6},\,10^{7}\, M_{\odot}$, and secondary mass $\mu = 50\, M_{\odot}$, with each source normalized to $\mathrm{SNR}=30$. The dimensionless primary spin is $\tilde{S} = 0.25$ and the eccentricity at the LSO is $e_\mathrm{LSO} = 0.1$. The orbital angles at the LSO are $\tilde{\gamma}_\mathrm{LSO} = \pi/2$, and $\alpha_\mathrm{LSO} = \Phi_\mathrm{LSO} = 0$, with inclination $\lambda = \pi/3$. The sky-location and spin-orientation angles are: $(\theta_S, \phi_S, \theta_K, \phi_K) = \left(2\pi/3, 5\pi/3, \pi/2, 0\right)$.
}
\centering
\small
\renewcommand{\arraystretch}{1.15}
\begin{tabular}{lccc}
\toprule
$M$ & $10^{5}M_{\odot}$ & $10^{6}M_{\odot}$ & $10^7 M_\odot$ \\
\midrule
$\DeltaESB$ & $5.98\times10^{-1}$ & $4.85\times10^{-1}$ & $3.16\times10^{0}$ \\
$\DeltaASB$ & $1.34\times10^{-3}$ & $7.42\times10^{-3}$ & $7.64\times10^{-2}$ \\
\bottomrule
\end{tabular}
\label{tablesconstraints}
\end{table}

In Table~\ref{tablesconstraints} we present the LISA measurement accuracies for ESB and ASB across EMRI primary masses, each normalized to $\mathrm{SNR}=30$. The axial constraint (ASB) is tightest for the lightest primaries, which accumulate the most strong-field cycles, reaching $\DeltaASB\simeq1.3\times10^{-3}$ at $M=10^5M_\odot$, and degrades toward $M=10^7M_\odot$ as the inspiral leaves the most sensitive part of the LISA band. The equatorial constraint (ESB), carried by the higher-order axisymmetric mass octupole, remains weak throughout, $\DeltaESB=\mathcal{O}(10^{-1}$--$1)$, and is effectively unconstrained at $M=10^7M_\odot$. We restrict the analysis to $M\ge10^5M_\odot$. At $M=10^4M_\odot$, the system is an intermediate-mass-ratio inspiral, where spin-orbit and spin-spin couplings not included here become relevant.

Figure~\ref{fig:q_vs_e} shows that the constraints on ASB are both tighter and more robust than those on ESB over a wide range of the eccentricity as measured at the LSO, $e_\mathrm{LSO}$ (the initial eccentricity in our simulations is in general significantly larger). 
This difference arises because ESB enters through a higher-order multipole moment, the axisymmetric mass octupole, which is more difficult to measure independently in EMRI observations than the leading quadrupolar contributions governing ASB. More precisely, within the mass-multipole sector that we consider, equatorial symmetry is broken only at octupole order, since the mass quadrupole is equatorially symmetric, whereas axial symmetry is already broken at quadrupole order by $Q_+$.

~

\noindent\textbf{Conclusions and discussion}. 
The results of our study demonstrate the strong potential of EMRI observations with LISA~\cite{LISA:2017pwj,Colpi:2024xhw} to probe the geometric structure of the most compact objects in the Universe. 
By measuring the subtle imprints of the multipolar structure of the EMRI primary in the gravitational waveform, LISA will enable precision tests of whether the most compact astrophysical objects are well described by the Kerr solution of general relativity. 
Our analysis shows that, for representative EMRI systems with SNR $\sim 30$, LISA can constrain deviations associated with axial-symmetry breaking at the $\sim10^{-2}$--$10^{-3}$ level and equatorial-symmetry breaking in the mass multipole sector at the $\sim10^{-1}$ level.
Moreover, these constraints improve significantly for primary masses in the range $M\sim10^{5}$-$10^{6}M_{\odot}$, where EMRI signals are optimally located within the LISA sensitivity band.
We also find that ASB is consistently measured with higher precision and robustness than ESB, reflecting the fact that it enters at quadrupolar order, whereas ESB is associated with higher-order multipolar structure.
These results establish concrete observational targets for testing strong-field deviations from Kerr and, in particular, for exotic compact objects such as fuzzballs.

Our framework (see~\cite{Muguruza:2026hqn} for details) allows us to incorporate a general multipolar structure for the EMRI primary that captures the leading dynamical effects that leave measurable imprints on the gravitational waveforms. Despite its simplified nature, and although it is based on underlying Newtonian dynamics as the AK model, our approach yields parameter-estimation forecasts in agreement with previous studies that include relativistic orbital effects. This supports the robustness of our conclusions and shows that LISA will be sensitive to deviations in the multipolar structure of ultracompact objects.

These results motivate extensions of our framework along several directions.  
First of all, we can specialize the analysis to exotic compact objects for which we have detailed theoretical knowledge of their dynamics.  This is still challenging, as such objects appear in complex theories and, as in the fuzzball case, they are expected to lack the symmetries that can simplify the problem. Nevertheless, some progress has been made in related settings, including EMRIs with boson star primaries~\cite{Macedo:2013jja,Destounis:2023khj} and systems involving point scalar charges orbiting topological stars~\cite{Bianchi:2024rod,DiRusso:2025lip,Melis_2025}.  
A complementary research direction is to incorporate relativistic orbital effects in the EMRI dynamics, by modeling the motion as geodesics of a perturbed stationary and axisymmetric spacetime describing the fuzzball model. Although this would provide a more faithful description of the strong-field regime, it is very challenging due to the limited knowledge of the analytic form of the metric of these exotic compact objects. However, some progress may be achieved by working in the slow-rotation (low spin) approximation.

Another promising direction is to extend our analysis to IMRIs, characterized by mass ratios in the range $10^{-4}\lesssim q\lesssim 10^{-2}$ (see~\cite{LISAConsortiumWaveformWorkingGroup:2023arg}).  These systems may arise if IMBHs exist either in globular clusters or as companions to massive BHs in galactic nuclei~\cite{LISA:2022yao}. Although IMRIs typically produce shorter-duration signals than EMRIs, their dynamics is more intricate, with spin-orbit and spin-spin couplings playing a more relevant role, with the spin of the secondary compact object contributing non-negligibly. These effects will certainly introduce additional structure in the gravitational waveform, which may enhance parameter-estimation accuracy and lead to more stringent constraints than those obtained in the present study.

\textbf{Acknowledgments}. P.F.M. and C.F.S. are supported by Contracts No. PID2022-137674NB-I00/AEI/10.13039/501100011033 (Spanish Ministry of Science and Innovation) and No. 2017-SGR-1469 (AGAUR, Generalitat de Catalunya). This work was also partly supported by the Spanish program Unidad de Excelencia María de Maeztu CEX2020-001058-M, financed by MCIN/AEI/10.13039/501100011033, and by the MaX-CSIC Excellence Award MaX4-SOMMA-ICE. P.F.M. also acknowledges financial support from Spanish Grant No. PRE2022-101913 funded by Spanish Ministry of Science and Innovation.

\textbf{Data availability}. The data that support the findings of this article are not publicly available. The data are available from the authors upon reasonable request.


%

\end{document}